\documentclass[aps,prd, twocolumn,superscriptaddress,nofootinbib,preprintnumbers]{revtex4-1}
\usepackage{amsmath}
\usepackage{graphicx}
\usepackage{subfigure}
\usepackage{amssymb}
\usepackage{xcolor}
\usepackage{multirow}
\usepackage{cancel}
\usepackage{color}
\usepackage{ulem}
\usepackage{listings}
\usepackage{xcolor}
\usepackage{float}
\usepackage{slashed}

\usepackage{tikz}
\usetikzlibrary{arrows,shapes}
\usetikzlibrary{trees}
\usetikzlibrary{matrix,arrows} 				
\usetikzlibrary{positioning}				
\usetikzlibrary{calc,through}				
\usetikzlibrary{decorations.pathreplacing}  
\usepackage{pgffor}							

\usetikzlibrary{decorations.pathmorphing}	
\usetikzlibrary{decorations.markings}
\tikzset{
    vector/.style={decorate, decoration={snake}, draw},
	provector/.style={decorate, decoration={snake,amplitude=2.5pt}, draw},
	antivector/.style={decorate, decoration={snake,amplitude=-2.5pt}, draw},
    fermion/.style={draw=black, postaction={decorate},
        decoration={markings,mark=at position .55 with {\arrow[draw=black]{>}}}},
    fermionbar/.style={draw=black, postaction={decorate},
        decoration={markings,mark=at position .55 with {\arrow[draw=black]{<}}}},
    fermionnoarrow/.style={draw=black},
    gluon/.style={decorate, draw=black,
        decoration={coil,amplitude=4pt, segment length=5pt}},
    scalar/.style={dashed,draw=black, postaction={decorate},
        decoration={markings,mark=at position .55 with {\arrow[draw=black]{>}}}},
    scalarbar/.style={dashed,draw=black, postaction={decorate},
        decoration={markings,mark=at position .55 with {\arrow[draw=black]{<}}}},
    scalarnoarrow/.style={dashed,draw=black},
    electron/.style={draw=black, postaction={decorate},
        decoration={markings,mark=at position .55 with {\arrow[draw=black]{>}}}},
	bigvector/.style={decorate, decoration={snake,amplitude=4pt}, draw},
}

\tikzstyle{block} = [draw, rectangle, 
    minimum height=3em, minimum width=6em]

\usepackage[colorlinks,citecolor=blue]{hyperref}
\usepackage{amsmath}
\usepackage{wrapfig}

\newcommand{\be}{\begin{equation}}
\newcommand{\ee}{\end{equation}}
\newcommand{\beq}{\begin{equation}}
\newcommand{\eeq}{\end{equation}}
\newcommand{\bea}{\begin{eqnarray}}
\newcommand{\eea}{\end{eqnarray}}
\newcommand{\besp}{\begin{equation}\begin{split}}
\newcommand{\eesp}{\end{split}\end{equation}}

\newcommand{\Eq}[1]{Eq.~(\ref{#1})}

\newcommand{\Dfbd}{\mathord{\buildrel{\lower3pt\hbox{$\scriptscriptstyle\leftrightarrow$}}\over {D}_{\mu}}}

\def\0{\textbf{0}}
\def\1{\textbf{1}}
\def\2{\textbf{2}}
\def\3{\textbf{3}}
\def\4{\textbf{4}}
\def\5{\textbf{5}}
\def\6{\textbf{6}}
\def\7{\textbf{7}}
\def\8{\textbf{8}}
\def\9{\textbf{9}}

\def\d{\text{d}}

\begin{document}

\title{Self-interacting dark matter implied by nano-Hertz gravitational waves}

\author{Chengcheng Han}
\email{hanchch@mail.sysu.edu.cn}
\affiliation{School of Physics, Sun Yat-sen University, Guangzhou 510275, P. R. China}

\author{Ke-Pan Xie}
\email{kpxie@buaa.edu.cn (corresponding author)}
\affiliation{School of Physics, Beihang University, Beijing 100191, P. R. China}

\author{Jin Min Yang}
\email{jmyang@itp.ac.cn}
\affiliation{CAS Key Laboratory of Theoretical Physics, Institute of Theoretical Physics, Chinese Academy of Sciences, Beijing 100190, P. R. China}
\affiliation{School of Physical Sciences, University of Chinese Academy of Sciences, Beijing 100049, P. R. China}

\author{Mengchao Zhang}
\email{mczhang@jnu.edu.cn (corresponding author)}
\affiliation{Department of Physics and Siyuan Laboratory, Jinan University, Guangzhou 510632, P.R. China}

\begin{abstract}

The self-interacting dark matter (SIDM) paradigm offers a potential solution to the small-scale structure problems faced by collision-less cold dark matter. This framework incorporates self-interactions among dark matter particles, typically mediated by a particle with a MeV-scale mass. Recent evidences of nano-Hertz gravitational waves from pulsar timing arrays (PTAs) such as NANOGrav, CPTA, EPTA, and PPTA suggest the occurrence of a first-order phase transition (FOPT) at a MeV-scale temperature. Considering the close proximity between these two scales, we propose that the mediator mass in the SIDM model originates from the spontaneous breaking of a $U(1)'$ symmetry, which is driven by the FOPT indicated by PTA data. Consequently, the alignment of these two scales is believed to be deeply connected by the same underlying physics. By extensively exploring the parameter space, remarkably, we find that the parameter space favored by SIDM just provides an explanation for the PTA data. 

\end{abstract}

\maketitle

\newpage
\section{Introduction}

The widely accepted cold dark matter (CDM) model successfully explains the Universe's structure and evolution. 
However, it faces challenges when addressing small-scale structure problems~\cite{Moore:1999gc, Moore:1994yx, Flores:1994gz, Oman:2015xda, Moore:1999nt, Klypin:1999uc, Boylan-Kolchin:2011lmk, Boylan-Kolchin:2011qkt}. 
These difficulties arise in understanding the behavior and distribution of dark matter (DM) within galaxies and galaxy clusters. To address these challenges, the self-interacting dark matter (SIDM) paradigm has emerged. 
It suggests that dark matter particles can interact through short-range forces mediated by a new particle called ``dark photon'' or ``dark mediator,'' typically with a mass at the MeV range. Validating the SIDM paradigm requires crucial searches for evidence of the existence of the dark sector.

Recently, the NANOGrav, CPTA, EPTA and PPTA collaborations have presented new observations of stochastic gravitational waves (GWs) using pulsar timing arrays (PTAs)~\cite{NANOGrav:2023gor, Xu:2023wog, Antoniadis:2023ott, Reardon:2023gzh}. In particular, the NANOGrav~\cite{NANOGrav:2023gor} and CPTA~\cite{Xu:2023wog} collaborations report a signal significance $\sim4\sigma$ for the Hellings-Downs correlation curve. These observations provide compelling evidence for the presence of stochastic GWs with a peak frequency around $10^{-8}$ Hz. While the standard interpretation has been inspiraling supermassive black hole binaries (SMBHBs), alternative explanations such as a first-order phase transition (FOPT) remain viable. It is known that a GW signal at $10^{-8}$ Hz implies a FOPT at the MeV scale, and a Bayesian analysis of the NANOGrav data even favors the FOPT model over the baseline SMBHB model~\cite{NANOGrav:2023hvm}. Therefore, this observation potentially shows the first evidence of signals from the early Universe prior to the Big Bang nucleosynthesis (BBN) and Cosmic Microwave Background (CMB). Studies on the theoretical models to explain the previous NANOGrav data can be found in Refs.~\cite{Ellis:2020ena, Blasi:2020mfx, Addazi:2020zcj, Buchmuller:2020lbh, Addazi:2020zcj, Vaskonen:2020lbd, DeLuca:2020agl, Domenech:2020ers, Addazi:2020zcj, Kohri:2020qqd, Bian:2020urb, Nakai:2020oit,Ratzinger:2020koh, Borah:2021ftr, Freese:2022qrl, Bringmann:2023opz, Madge:2023cak, Kobakhidze:2017mru, Arunasalam:2017ajm,Chiang:2020aui,Ashoorioon:2022raz,Nakai:2020oit,Bigazzi:2020avc,Freese:2023fcr}.

The remarkable proximity between the scale of mediating DM self-interaction and that of the FOPT indicated by the PTA data suggests a profound connection between them. We propose that the SIDM mediator mass originates from the spontaneous symmetry breaking of a new gauge sector, which occurs through a strong FOPT in the early Universe. Consequently, the alignment of these two scales signifies the presence of a unified underlying physics that governs the dynamics within the dark sector.

In this work, we present a concise and comprehensive model within the framework of SIDM to explain the recently reported PTA data, specifically focusing on the NANOGrav and CPTA observations. By thoroughly investigating our proposed model, we explore and identify a parameter space that can successfully account for the observed GW signals while addressing the small-scale structure problems. This model holds promise for experimental testing in the near future. 

The article is organized as follows: Section \ref{model} provides an introduction to our model and explores the physics associated with SIDM. Section \ref{phase1} focuses on the calculation of the FOPT and the corresponding GW signals. In Section \ref{result}, we present our numerical results, showcasing the viable parameter space, and engage in further discussions. Finally, we conclude in Section \ref{conclu}.

\section{Self-interacting dark matter}
\label{model}

Small-scale problems are series of discrepancies between astrophysical observations and the simulation of collision-less CDM at the scale smaller than $\mathcal{O}(1)$ Mpc.
For example, the diversity of inner rotation curves of spiral galaxies is difficult to be explained by CDM~\cite{Oman:2015xda}. 
However, if there is an elastic scattering process between DM particles, the inner region of DM halo will be heated up and avoid being too dense due to thermalization, then  the diversity problem can be alleviated~\cite{Kamada:2016euw,Creasey:2017qxc,Ren:2018jpt}.

Overall, for dwarf galaxies or galaxies with a DM average velocity $ \langle v \rangle$ around $10-200$ km/s, the cross section $\sigma/m_{\text{DM}}$ needs to be within the range $\mathcal{O}(1)-\mathcal{O}(10)$ cm$^{2}$/g ~\cite{Zavala:2012us,Elbert:2014bma}. However, for galaxy clusters where $ \langle v \rangle \sim 2000$ km/s, current fit result favors $\sigma/m_{\text{DM}} \sim 0.2 $ cm$^{2}$/g~\cite{Sagunski:2020spe}. It has been shown that a velocity-dependent $\sigma/m_{\text{DM}}$ can solve small-scale problems at different systems~\cite{Kochanek:2000pi,Andrade:2020lqq,Elbert:2016dbb,Fry:2015rta,Yoshida:2000bx,Moore:2000fp,Zavala:2012us,Elbert:2014bma,Rocha:2012jg,Peter:2012jh,Kaplinghat:2015aga,Tulin:2017ara,Vogelsberger:2014pda,Dooley:2016ajo,Dave:2000ar,Robertson:2018anx,Spergel:1999mh,Colin:2002nk,Vogelsberger:2012ku,Harvey:2015hha,Burkert:2000di,Sagunski:2020spe,Yoshida:2000uw,Yang:2023jwn}.
Such a velocity-dependent cross section can be easily induced via a light mediator between DM particles~\cite{Tulin:2017ara}. 

In this work, we propose a model that the mediator $A'$ is the gauge boson of a spontaneous broken dark $U(1)'$ symmetry. The breaking of this $U(1)'$ is through a FOPT in the elarly Universe, which generates the $10^{-8}$ Hz GWs, as we will show later. The corresponding Lagrangian  for the dark sector is given by 
\begin{multline}\label{Lagrangian}
\mathcal{L}_{U(1)'} = \bar{\chi}( i \slashed{D} - m_{\chi} ) \chi - \frac{1}{4} F'_{\mu\nu}  F'^{\mu\nu}\\  +(D_{\mu} S)^{\dagger} D^{\mu} S  - V(S)~, 
\end{multline} 
where $\chi$ is the Dirac DM candidate, and $S=(\phi+i\eta)/\sqrt{2}$ is the dark Higgs field whose potential is
\begin{eqnarray}\label{V0}
V(S) = - \mu^2 S^{\dagger} S + {\lambda} \left( S^{\dagger}S \right)^2~.
\end{eqnarray} 
$D_{\mu} \equiv \partial_{\mu} + i Q_{i} g' A'_{\mu} $ with $g'$ being the dark gauge coupling.
For $U(1)'$ charge assignment we choose $\{Q_{\chi},Q_{S}\} = \{+3/2,+1\}$~\footnote{The reason for such a charge assignment will be given in the supplemental material.}. The Mexican hat shape of potential (\ref{V0}) generates a nonzero vacuum expectation value $v_s=\mu/\sqrt{\lambda}$, breaking the $U(1)'$ spontaneously, resulting in a massive $A'$ with $m_{A'}=g'v_s$ and dark Higgs boson $\phi$ with $m_\phi=\sqrt{2}\mu$. We emphasize that the simple Lagrangian we presented here is primarily focused on the phase transition and dark matter self-interaction. However, to avoid the limit of the BBN and CMB, an extension of the model is needed and we will give the details in the supplemental material.    

Previous studies show that a DM candidate with a mass $m_\chi \sim \mathcal{O}(10)$ GeV and a mediator with a mass $m_{A'} \sim \mathcal{O}(10)$ MeV can well fit the DM data from dwarf galaxies to galaxy clusters~\cite{Boddy:2014qxa,Feng:2009hw,Feng:2009mn,Kamada:2018kmi,Kamada:2018zxi,Kamada:2019gpp,Kamada:2019jch,Kamada:2020buc,Kang:2015aqa,Ko:2014bka,Ko:2014nha,Schutz:2014nka,Tulin:2012wi,Tulin:2013teo,Wang:2016lvj,Wang:2014kja,Wang:2022lxn,Wang:2022akn,Chen:2023rrl}. In our model, for $v_s \sim \mathcal O$(10) MeV and $g'\sim \mathcal{O}$(1), a mediator mass $m_{A'}$ around 10 MeV can be easily generated. On the other hand, the scale of $v_s$ also indicates the phase transition occurs at a temperature around MeV.

To fit the small-scale data, one needs to calculate the thermal averaged DM scattering cross section $ \overline{ \sigma }$~\cite{Colquhoun:2020adl}. Here $ \overline{ \sigma }$ is given by
\begin{equation}
    \overline{ \sigma } \equiv \frac{\langle \sigma_{V} v^3_{\text{rel}}  \rangle }{ 24 \sqrt{\pi} v^3_0}~;\quad \sigma_V \equiv \int \text{d} \Omega \sin^2\theta \frac{\text{d}\sigma}{\text{d} \Omega}~,
\end{equation}
where $\sigma_V$ is the viscosity cross section~\cite{Tulin:2013teo,Cline:2013pca,Yang:2022hkm} and $v_0$ is the velocity dispersion. Analytic formulae of DM scattering cross section at different parameter regions corresponding to different coupling strength and kinetic energy have been given in the literature~\cite{Tulin:2013teo,Feng:2009hw,Buckley:2009in,Kahlhoefer:2017umn,Khrapak:2003kjw,Cyr-Racine:2015ihg,Colquhoun:2020adl,Girmohanta:2022dog,Girmohanta:2022izb}, and implemented in public package {\tt CLASSICS}~\cite{Colquhoun:2020adl}, which is used in our calculation. The fit result will be given in Section~\ref{result}.

It should be pointed out that the $A'$-induced Sommerfeld effect~\cite{Sommerfeld:1931qaf} will largely enhance the $\chi\bar{\chi}$ annihilation during the recombination period,  leading to severe constraints from CMB~\cite{Galli:2009zc,Slatyer:2009yq,Bringmann:2016din}. 
However, such limit can be easily evaded in the asymmetric DM scenario in which the dark sector has a nonzero $Y_{\Delta\chi}=(n_\chi-n_{\bar\chi})/s$, similar to the baryon asymmetry in the visible sector~\cite{Chen:2023rrl, Mohapatra:2001sx,Frandsen:2010yj,Petraki:2014uza,Dessert:2018khu,Dutta:2022knf,Heeck:2022znj,Hall:2019rld,Hall:2021zsk,Tsai:2020vpi,Chen:2023rrl}. The detail of the generation of the dark asymmetry and correct dark matter relic can be found in supplemental material Sec.~II. Here we also assume the chemical potential in the dark sector is small enough to not affect the FOPT of the dark sector.

\section{First-order phase transition from dark sector}
\label{phase1}

Shortly after the inflationary reheating, the Universe enters the radiation era with high temperature and density. We assume the dark and visible sectors are in thermal equilibrium (see Section~\ref{result}). The scalar potential receives thermal corrections and becomes temperature dependent. The one-loop finite temperature potential reads
\begin{multline}\label{VT}
V_T(\phi,T)=V_0(\phi)+V_1(\phi)+\delta V(\phi)\\    +V_{T1}(\phi,T)+V_{\rm daisy}(\phi,T)~,
\end{multline}
where $V_0$ is Eq.~(\ref{V0}) with $S\to\phi/\sqrt{2}$, $V_1$ is the Coleman-Weinberg potential, $\delta V$ is the counter term, $V_{T1}$ is the thermal correction, and $V_{\rm daisy}$ is the daisy resummation. The complete form of $V_T(\phi,T)$ is given in supplemental material Sec.~I with the comparison with Ref.~\cite{Niedermann:2021vgd}.

At zero temperature, the $U(1)'$ symmetry is spontaneously broken. However, at high temperatures in the early Universe, thermal corrections can restore the $U(1)'$ symmetry. This is evident from $V_T(\phi,T)\approx(-\mu^2+g'^2T^2/4)\phi^2/2$ at $\phi\sim0$, where a positive mass square term arises due to sufficiently high temperatures~\cite{Dolan:1973qd}. Initially the Universe is at  $\phi=0$. As it cools down and the potential shape changes, it transitions to the vacuum state $\phi\neq0$. This transition can occur smoothly through the rolling of the $\phi$ field. However, in certain parameter ranges, a potential barrier induced by the $A'$ field separates the two vacua, leading to a discontinuous quantum tunneling process known as a FOPT that proceeds via vacuum bubble nucleation and expansion.

\begin{table*}[!htbp]
\begin{tabular}{ c c c c c c c c}
\hline
\hline
Benchmark point &\ $m_\chi$ [GeV]  \ &\ $m_{A'}$ [MeV]  \ &\ $m_{\phi}$ [MeV]  \ &\ $g'$  \  &\ $T_n$ [MeV] \ &\ ${\beta}/{H_n}$ \ &\ $\alpha$  \\ 
\hline
\hline
BP1  & \  32.0 \  & \ 21.7 \  & \ 4.40 \  & \ 0.960 \  & \ 5.81 \  & \ 15.8 \  & \ 0.326    \\
\hline
BP2  & \ 36.0 \  & \ 19.8 \  & \ 4.01 \  & \ 0.985 \  & \ 5.00 \  & \ 20.0 \  & \ 0.356     \\
\hline
BP3  & \ 31.0 \  & \ 22.1 \  & \ 4.10 \  & \ 0.921 \  & \ 4.72 \  & \ 12.0 \  & \ 0.542    \\
\hline
BP4  & \ 32.0 \  & \ 22.3 \  & \ 4.51 \  & \ 0.990 \  & \ 5.56 \  & \ 28.8 \  & \ 0.363    \\
\hline
\end{tabular}
\caption{ Relevant parameters for the  benchmark points of our model.  }
\label{bp}
\end{table*}

To quantitatively calculate the FOPT dynamics, one needs to solve the $O(3)$ symmetric bounce solution for $V_T$ and derive the classical action $S_3$, and decay probability per unit volume is then~\cite{Linde:1981zj}
\begin{equation}
    \Gamma(T)\approx T^4\left(\frac{S_3}{2\pi T}\right)^{3/2}e^{-S_3/T}.
\end{equation}
The nucleation of bubbles containing the true vacuum occurs at a temperature $T_n$ when the decay probability within a Hubble volume and a Hubble time, given by $\Gamma(T_n)H^{-4}(T_n)$, reaches $\mathcal{O}(1)$. Note that bubble nucleation itself does not guarantee the successful completion of a FOPT. A more rigorous criterion of FOPT is the existence of percolation temperature $T_p$ at which the true vacuum bubbles form an infinite connected cluster, and the volume of space occupied by the true vacuum keeps increasing~\cite{Ellis:2018mja}. However, for a mild FOPT (as considered here), which ensures completion once nucleation occurs, we adopt the nucleation condition as a practical criterion and rewrite it as 
\begin{equation}\label{FOPT_criterion}
\frac{S_3}{T_n}\approx4\log\left(\frac{1}{4\pi}\sqrt{\frac{45}{\pi g_*(T_n)}}\frac{M_{\rm Pl}}{T_n}\right),
\end{equation}
where $M_{\rm Pl}=1.22\times10^{19}$ GeV is the Planck scale, and $g_*$ is the number of relativistic degrees of freedom. We focus on $T_n\sim$ MeV, where $g_*=10.75$ and the right-hand-side of Eq.~(\ref{FOPT_criterion}) is around $190$.

During a FOPT, the collision of bubbles, the motion of sound waves and turbulence in the plasma generate stochastic GWs~\cite{Athron:2023xlk}. Typically, the bubble walls reach a terminal velocity due to the friction force exerted by the plasma particles, and most of the released vacuum energy goes to surrounding plasma. As a consequence, the main GW sources are sound wave and turbulence, and the former usually dominates~\cite{Ellis:2019oqb}. The GW spectrum today, defined as
\begin{equation}
    \Omega_{\rm GW}(f)=\frac{1}{\rho_c}\frac{\text{d}\rho_{\rm GW}}{\text{d}\log f},
\end{equation}
with $\rho_{\rm GW}$ and $\rho_c$ being the GW and current Universe energy density, respectively, can be written as numerical functions of the FOPT parameters $\{\alpha,\beta/H_*,T_*,v_w\}$~\cite{Caprini:2015zlo}, where $\alpha$ is the ratio of latent heat to the radiation energy density, $\beta/H_*$ is the inverse ratio of the FOPT duration to the Hubble time scale, $T_*$ is the temperature and $v_w$ is the bubble velocity. The sound wave contribution is ~\cite{Hindmarsh:2015qta}
\small 
\begin{multline}\label{sw}
    \Omega_{\rm sw}(f)h^2=5.71\times10^{-8}v_w\left(\frac{\kappa_V\alpha}{1+\alpha}\right)^2\left(\frac{\beta/H_*}{100}\right)^{-1} \\
     \times \left(\frac{g_*}{10}\right)^{-1/3}\left(\frac{f}{f_{\rm sw}}\right)^3\left(\frac{7}{4+3(f/f_{\rm sw})^2}\right)^{7/2},
\end{multline}
\normalsize 
where $\kappa_V$ is the fraction of vacuum energy that is released to bulk motion, and
\begin{multline}
f_{\rm sw}=1.3\times10^{-8}~{\rm Hz}\\
\times\frac{1}{v_w}\left(\frac{\beta/H_*}{100}\right) 
\left(\frac{T_*}{\rm MeV}\right)\left(\frac{g_*}{10}\right)^{1/6},
\end{multline}
is the peak frequency, which implies  $f_{\rm sw}\sim10^{-8}$ Hz  for $T_*\sim$ MeV.

To obtain the GW spectrum, we use the package {\tt CosmoTransitions}~\cite{Wainwright:2011kj} to resolve the bounce equation and derive $S_3/T$ and hence derive $T_n$ via \Eq{FOPT_criterion}, and analyze the hydrodynamic motion of the plasma using the bag model to determine the FOPT energy budget~\cite{Espinosa:2010hh}. Numerical simulations from previous studies are then utilized to calculate $\Omega_{\rm GW}(f)h^2$ contributions from sound waves and turbulence, where $T_*=T_n$ and $H_*=H(T_n)$ are adopted. According to \Eq{sw}, the GW signal strength has a mild positive correlation with the bubble expansion velocity $v_w$.
The velocity is determined by the competition between the outward vacuum pressure and the friction force induced by the particles interacting with the bubble wall, and numerical simulations show that the electroweak FOPTs in the singlet scalar extended SM typically exhibit a wall velocity of $v_w\gtrsim0.6$~\cite{Laurent:2022jrs,Ellis:2022lft}. In our scenario, the dark $U(1)'$ FOPT only features the friction force from the $A'$ boson, thus we expect $v_w$ can be larger. As a benchmark, we plot the envelop of GW spectra for $v_w\in[0.6,1]$. We account for the finite lifetime of sound waves through an additional suppression factor $H \tau_{\rm sw}\leqslant1$~\cite{Guo:2020grp}. While turbulence usually has a sub-leading contribution, it modifies the tail of the sound wave GW signal, thus we include its effect in our analysis using results from Ref.~\cite{Binetruy:2012ze}.

\section{Results and Discussion}
\label{result}

\begin{figure*}
\begin{center}
\includegraphics[width=6.8in]{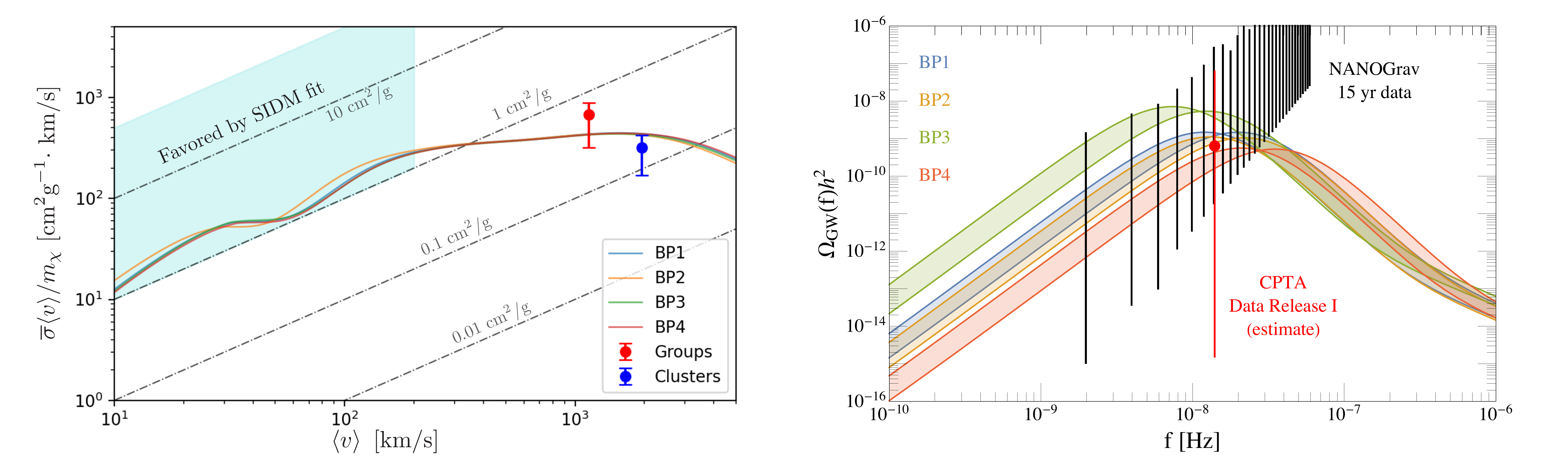}
\caption{Velocity-dependent DM elastic scattering cross sections (left) and GW spectra (right) for our benchmark points. See the text for details.}
\label{SIGW}
\end{center}
\end{figure*}

Using the established framework, we conducted numerical calculations to determine the parameter space necessary to address both the small-scale problems and explain the PTA data. By systematically exploring this parameter space, we identified specific benchmark points (BPs) that simultaneously reconcile the PTA data and resolve the small-scale problems, as summarized in Table~\ref{bp}. In the left and right panels of Fig.~\ref{SIGW}, we present the velocity-dependent elastic cross-sections of DM particles and the corresponding GW spectra derived from the BPs, respectively. 
As we explained in Section~\ref{model}, in the velocity region $10 -200$ km/s, we require the cross section $\overline{\sigma}/m_\chi$ to be within range $\mathcal{O}(1)-\mathcal{O}(10)$ cm$^{2}$/g ~\cite{Zavala:2012us,Elbert:2014bma}. 
For galaxy groups where $\langle v \rangle \approx 1150 $ km/s and galaxy clusters where $\langle v \rangle \approx 1900 $ km/s, we use the result given in~\cite{Sagunski:2020spe} which requires $\overline{\sigma}/m_\chi = 0.5 \pm 0.2$ cm$^{2}$/g and $\overline{\sigma}/m_\chi = 0.19 \pm 0.09$ cm$^{2}$/g, respectively. 
The NANOGrav data are from the collaboration result~\cite{NANOGrav:2023gor}, while the CPTA data point is converted from the best fit point of $f=14$ nHz and ${\rm lg}A=-14.4^{+1.0}_{-2.8}$~\cite{Xu:2023wog}. Remarkably, we find that the identified BPs not only satisfy the required cross section values to explain the small-scale problems but also accommodate the observational data from NANOGrav and CPTA.

Since the FOPT temperature is relatively low, a few discussions on other cosmological constraints are in order. As $T_n\gtrsim3$ MeV, the CMB and BBN constraints can be satisfied; for example, the effective number of neutrinos ($N_{\text{eff}}$) is essentially not affected by the FOPT reheating~\cite{Bai:2021ibt,Deng:2023seh}. The constraint from ultra-compact minihalo (UCMH) abundance is rather severe~\cite{Liu:2022lvz,Liu:2023hte}, but this issue can be relaxed by choosing a conservative value of the red shift of the last formation of UCHM at $z_c=1000$.

We further require the lifetime of $\phi$ to be short enough to avoid energy injection to the visible sector during the BBN period. For particles with mass larger than about 4 MeV and an $e^+ e^-$ decay final state, the lifetime should be shorter than 0.1 s, assuming the dark sector number of degrees of freedom at MeV is 1~\cite{Depta:2020zbh,Ibe:2021fed}. In our model, the decay width of $\phi$ is suppressed by its tiny mixing angle with the SM Higgs boson, which is constrained to be $|\theta|\lesssim10^{-5}$ by the Higgs exotic decay~\cite{CMS:2018yfx}, making the lifetime exceed 10 s. However, this bound can be avoided by extending the Higgs sector with an additional Higgs doublet, whose mixing with the $\phi$ can enhance the $\phi e^-e^+$ coupling, leading to a safely fast-decaying $\phi$. We give a detail of the model in supplemental material.  

\section{Conclusion}\label{conclu}

In this study, we established that a FOPT at the MeV scale, as inferred from the PTA data, provides a compelling mechanism for generating a mediator mass within the MeV range, which is crucial in addressing the small-scale problem within the framework of the SIDM. By considering various constraints, we identified the parameter space where the small-scale structure problem can be effectively resolved while simultaneously fitting the PTA data.


Our model has rich  phenomenology. Direct detection experiments ($\chi N \to \chi N$ mediated by $A'$) like PandaX-II~\cite{PandaX-II:2021lap} constrain the kinetic mixing between $A'$ and photons to $\epsilon\lesssim10^{-11}$ for $m_\chi \sim \mathcal{O}(10)$ GeV and $m_{A'} \sim \mathcal{O}(10)$ MeV. We note that the introduction of the additional Higgs would loop-induced kinetic mixing of the visible photon and dark photon. However, it has been found that the loop induced kinetic mixing of the visible photon and dark photon is highly suppressed~\cite{Koren:2019iuv}, which appears at five loops in our model. The combination of loop factors and coupling suppression provides sufficient suppression to maintain consistency with current limits. 
Furthermore, the very distinct gravothermal evolution of SIDM halo, which start with core formation-expansion and followed by core collapse~\cite{Balberg:2002ue,Koda:2011yb,Essig:2018pzq,Huo:2019yhk,Nishikawa:2019lsc,Sameie:2019zfo,Kahlhoefer:2019oyt,Turner:2020vlf,Zeng:2021ldo,Outmezguine:2022bhq,Yang:2022hkm,Yang:2022zkd,Yang:2023jwn,Nadler:2023nrd}, leave imprint on star formation history~\cite{Robles:2019mfq,Sameie:2021ang,Vargya:2021qza,Burger:2022cjo} and supermassive black holes seeding~\cite{Balberg:2001qg,Pollack:2014rja,Feng:2020kxv}.

In summary, the PTA data can help us better reveal the properties of SIDM by combing with those astronomy observations and shed light on the nature of DM. The upcoming experiments have the potential to detect various signals from our model, providing a complementary investigation of the new physics associated with nano-Hertz GWs.

\section*{Acknowledgements}

We would like to thank Jing Liu, Tao Xu and Daneng Yang for the very useful discussions. This work was supported by the National Natural Science Foundation of China (NNSFC) under grant Nos.  12105118, 11947118
11821505, 12075300 and 1233500, by Peng-Huan-Wu Theoretical Physics Innovation Center (12047503), and by the Key Research Program of the Chinese Academy of Sciences, grant No. XDPB15.  CH acknowledges support from the Sun Yat-Sen University Science Foundation and the Fundamental Research Funds for the Central Universities, Sun Yat-sen University under Grant No. 23qnpy58.

\appendix
\section{Detailed expressions of thermal potential}
\label{app:VT}

In this appendix we list the concrete expression of each term in Eq.~(4). The Coleman-Weinberg potential can be derived using the field-dependent mass as
\begin{eqnarray}
    V_1(\phi)&=&\frac{M_\phi^4(\phi)}{64\pi^2}\left(\log\frac{M_\phi^2(\phi)}{\mu_R^2}-\frac32\right)  \nonumber \\
     &&+3\frac{M_{A'}^4(\phi)}{64\pi^2}\left(\log\frac{M_{A'}^2(\phi)}{\mu_R^2}-\frac56\right)~,
\end{eqnarray} 
where $M_\phi^2(\phi)=-\mu^2+3\lambda\phi^2$, $M_\eta^2(\phi)=-\mu^2+\lambda\phi^2$, $M_{A'}(\phi)=g'\phi$, and $\mu_R$ is the renormalization scale which is adopted as 10 MeV. We have dropped the contribution from the Goldstone $\eta$ to avoid IR divergence~\cite{Espinosa:2011ax}. The counter term
\begin{equation}
    \delta V(\phi)=-\frac{\delta\mu^2}{2}\phi^2+\frac{\delta\lambda}{4}\phi^4~,
\end{equation}
is derived by the conditions
\begin{equation}
    \frac{\partial (V_1+\delta V)}{\partial\phi}\Big|_{\phi=v_s}=0,\quad
    \frac{\partial^2 (V_1+\delta V)}{\partial\phi^2}\Big|_{\phi=v_s}=0~,
\end{equation}
such that the zero temperature tree-level relations between $(\mu^2, \lambda)$ and $(v_s,m_{A'},m_\phi)$ still hold.

The one-loop thermal correction is
\small 
\begin{equation}
    V_{T1}(\phi,T)=\sum_{i=\phi,\eta,A'}\frac{n_iT^4}{2\pi^2}J_B\left(\frac{M_i^2(\phi)}{T^2}\right)~,
\end{equation}
\normalsize 
where $n_{\phi,\eta}=1$ and $n_{A'}=3$, and the Bose thermal integral is defined as
\begin{equation}
J_{B}(y)=\int_0^\infty x^2\d x\log(1- e^{-\sqrt{x^2+y}})~.
\end{equation}
Finally, the daisy resummation term is
\begin{equation}
  V_{\rm daisy}(\phi,T) = -\frac{g'^3 T}{12\pi} \left((\phi^2 + T^2)^{3/2} - \phi^3\right)~,
\end{equation}
where we only consider the dominant $A'$ contribution, as the parameter space of interest always has $g'^2\gg\lambda$. The above discussions define our finite temperature potential $V_T(\phi,T)$ in Eq.~(4).

\begin{figure}
\begin{center}
\includegraphics[width=9cm]{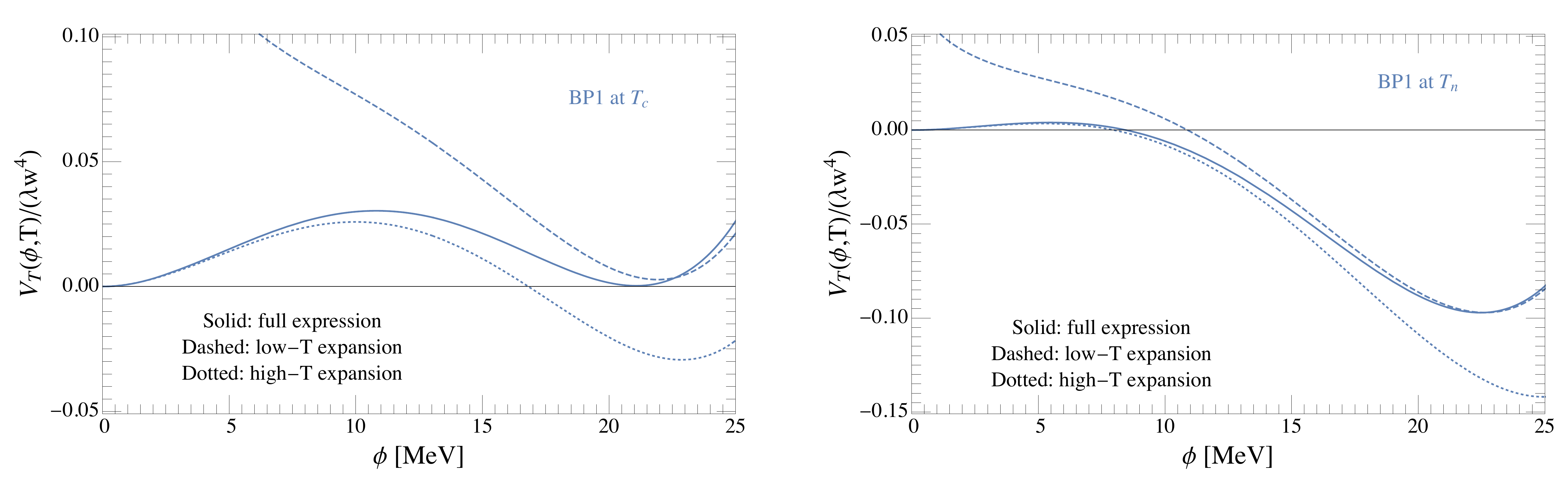}
\caption{The potential plots of BP1 at critical temperature $T_c$ (left) and nucleation temperature $T_n$ (right). The solid, dashed and dotted lines represent the complete one-loop, low temperature and high temperature expansions, respectively.}
\label{fig:potential}
\end{center}
\end{figure}

The potential of BP1 described in the main text are ploted in Fig.~\ref{fig:potential}, where the left panel represents $V_T$ at critical temperature $T_c$ at which two dengnerate vacua exist, while the right panel represents $V_T$ at nucleation temperature at which the bubbles start to emerge. The solid, dashed and dotted lines represent the complete one-loop, low temperature and high temperature expansions, respectively. We obtain $v_s/T_n\sim5$ for the BPs, as the observed GW signal requires a strong FOPT. This also implies that the low- or high-temperature expansions are not suited in this scenario, although they are good approximations at $\phi\sim v_s$ and small-$\phi$ limits, respectively~\cite{Niedermann:2021vgd}, as illustrated in Fig.~\ref{fig:potential}. Therefore, we include the complete one-loop expression to calculate the FOPT. The two-loop thermal correction is sub-dominant since we are considering the parameter space of $g'\sim1$ and $\lambda\sim0.01$, which does not satisfy $g'^4\ll\lambda\ll g'^2$~\cite{Arnold:1992rz}.


\section{Generation of DM asymmetry and CMB bound}\label{app:ADM}

In this appendix we detail how to generate the asymmetry in the dark sector, and how this asymmetry helps to escape the stringent limit from CMB. 
Similar analysis can be found in our previous study~\cite{Chen:2023rrl} and references therein. 

Firstly, an easy method to generate DM asymmetry is the CP violated and out-of-equilibrium decay of some heavy particle. 
Similar to the vanilla Leptogenesis mechanism, we introduce two flavors of neutral right handed neutrinos, $N_1$ and $N_2$, as the source of asymmetry. 
Corresponding Lagrangian is given by: 
\begin{eqnarray}
\mathcal{L}_\text{RHN} && = \frac{1}{2} \sum_{i=1,2} \bar{N}_i ( i \slashed{\partial} - M_{N_i} ) N^C_i  \nonumber \\
&& - \sum_{i=1,2} ( y_i \bar{N}_i \chi \zeta^{\dagger} + h.c. )~, 
\end{eqnarray} 
$N_1$ and $N_2$ can certainly couple to the SM leptons and generate the asymmetry in the visible sector. 
Here we assume couplings between $N_i$, SM leptons, and SM Higgs are negligible small and thus we can focus on the dark sector. 
We also introduce another dark scalar $\zeta$ with $U(1)'$ charge +3/2 to realize the darkgenesis. The new scalar $\zeta$ should have vanishing vacuum to avoid the majorana mass for the dark matter through the seesaw mechanism. If $\zeta$ has a mass much lighter than the dark matter mass, for example, 0.1 GeV, then the $\zeta$ becomes stable and becomes one of dark matter candidate. Since the relic number density of the $\zeta$ is same as the dark matter $\chi$, its abundance is much smaller than the fermion dark matter which contributes less than one percent of dark matter(this number can be reduced if the mass of $\zeta$ becomes even smaller). We do not expect it affect the dynamics of the main component of the dark matter and the discussion of the dark matter part would safely ignore it.


We choose $N_1$ to be the lighter one between $N_1$ and $N_2$. In the case where parameters $y_i$ have complex phases, the CP asymmetry generated in $N_1$ decay is given by:
\begin{eqnarray}
\epsilon &&\equiv \frac{\Gamma(N_1 \to \chi \zeta^{\dagger}) - \Gamma(N_1 \to \bar{\chi} \zeta)}{\Gamma(N_1 \to \chi \zeta^{\dagger}) + \Gamma(N_1 \to \bar{\chi} \zeta)}  \nonumber \\
&& \simeq -\frac{1}{16\pi} \frac{M_{N_1}}{M_{N_2}} \frac{ \text{Im} \left[ (y^{\ast}_2 y_1 )^2 \right] }{|y_1|^2}~.
\end{eqnarray}
To simplify expression, in the second line we consider $M_{N_2} \gg M_{N_1}$.

We assume $N_1$ is originally in the thermal bath with the standard model sector and dark matter sector. When the decay of $N_1$ is sufficiently out-of-equilibrium, the Yield of DM asymmetry $Y_{\Delta \chi}$ (defined as $Y_{\Delta \chi} \equiv Y_{\chi} - Y_{\bar{\chi}}$) is given by:
\begin{eqnarray}
Y_{\Delta \chi} = Y_{N_1} \times \epsilon \times \eta &\simeq& -\frac{Y_{N_1}}{16\pi} \frac{M_{N_1}}{M_{N_2}} \frac{ \text{Im} \left[ (y^{\ast}_2 y_1 )^2 \right] }{|y_1|^2}~,
\label{yield}
\end{eqnarray}
where $Y_{N_1}= \frac{n_{N_1}}{s} \simeq 0.004$ is the yield of $N_1$ before it decays, and wash-out factor $\eta$ can be around 1 depending on the parameters.

\begin{figure}
\begin{center}
\includegraphics[width=9cm]{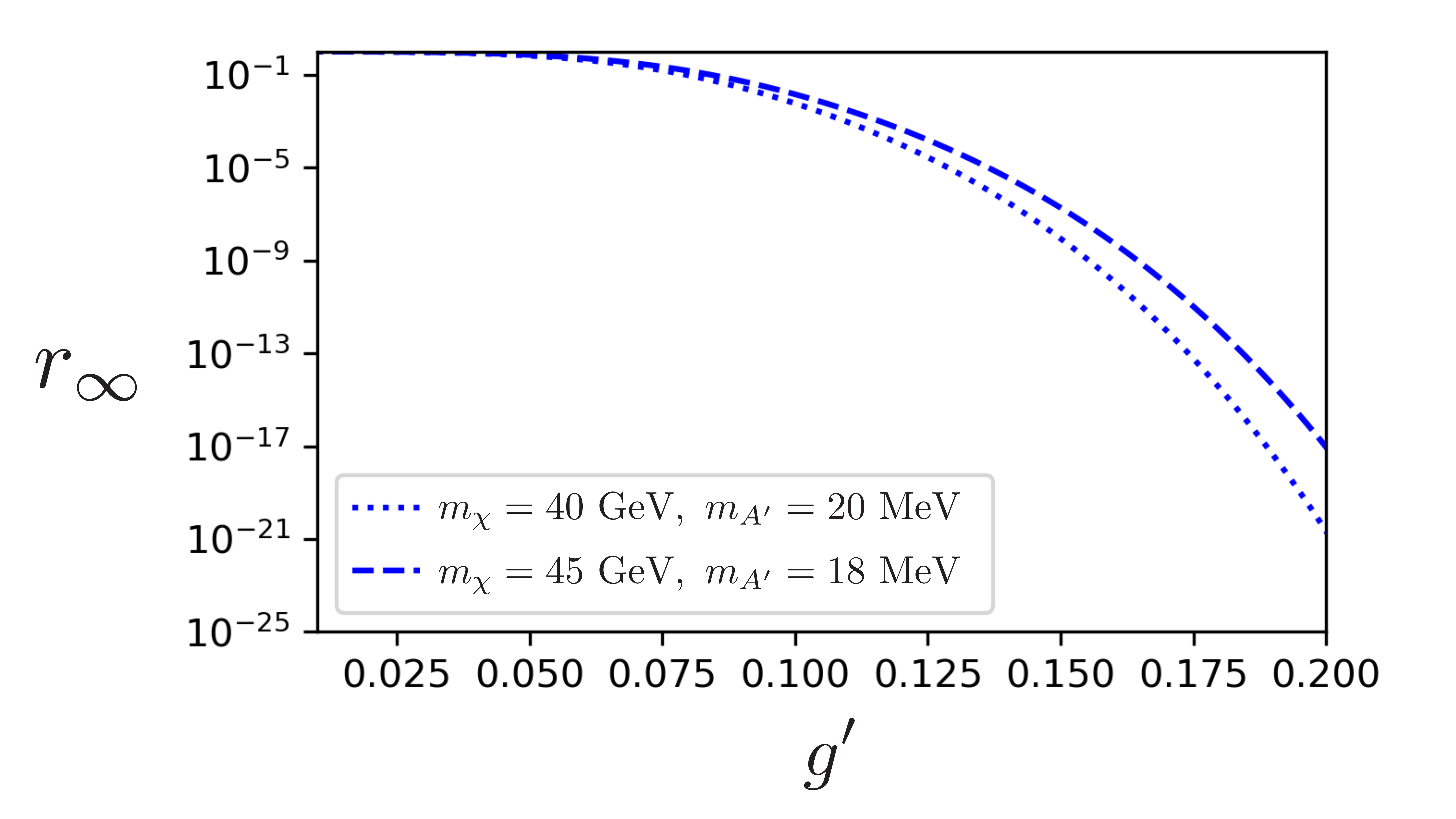}
\caption{$r_{\infty}$ as functions of $g'$ for different benchmark spectrum favored by small-scale data.}
\label{fig:r_value}
\end{center}
\end{figure}

Current DM relic density is: 
\begin{eqnarray}
\Omega_{\chi} h^2 \approx m_{\chi} Y_{\Delta\chi} s_0 / \rho_{\text{cr}} \approx Y_{\Delta\chi} \left( \frac{m_{\chi}}{\text{GeV}} \right) \times 2.72 \times 10^8~.  \nonumber \\
\end{eqnarray}
The correct value, i.e. $\Omega_{\chi} h^2 \approx 0.12$, can always be achieved via a combined tuning of parameters in Eq.~(\ref{yield}). 
For example, if $m_{\chi} = 40$ GeV, then setting: 
\begin{eqnarray}
&& M_{N_1} = 10^6\text{ GeV} \ , \ M_{N_2} = 10^7\text{ GeV} \ , \nonumber \\
&& y_2 = 0.05 + i 1.38\times 10^{-5} \ , \ y_1 = 10^{-6}~,
\end{eqnarray}
gives the correct relic density. 

It should be noticed that, after the out-of-equilibrium and CP broken decay of $N_1$, the asymmetry of $\zeta$ is also generated (i.e. $Y_{\Delta\zeta} = -Y_{\Delta\chi}$). 
Due to the $U(1)'$ charge assignment, only $(\zeta^\dagger \zeta)^2$ and $(\zeta^\dagger \zeta) (S^\dagger S)$ are allowed operators. Thus, in addition to $\chi$, $\zeta$ is also stable dark matter. 
Here we simply set $\zeta$ mass $m_\zeta$ to be around $\mathcal{O}(0.1)$ GeV scale.  
In this case, relic density is still dominated by $\chi$.  
$\mathcal{O}(0.1)$ GeV scale $\zeta$ will not affect BBN and $N_{\text{eff}}$, because $\zeta\zeta^\dagger \to A'A'$ annihilation will be negligible when the universe temperature is below MeV. 
The existence of $\zeta$ will not affect the discussion of DM small-scale structure in the main text, this is because the rate of momentum-changing inside DM halo is dominated by $\chi$~\cite{Petraki:2014uza}. 
Furthermore, dark atom composed by $\chi-\zeta^\dagger$ can not form due the MeV scale dark photon~\cite{Petraki:2014uza}. 
Thus we can conclude that in the following discussion of $\chi$, we can safely ignore the impact from $\zeta$.

Next we turn to the CMB bound. 
For later convenience, we label the ratio of DM and anti-DM number densities by $r$:
\begin{eqnarray}
r =  \frac{Y_{\bar{\chi}}}{Y_{{\chi}}}~, 
\end{eqnarray}
$r$ is certainly a function of time. 
Compared with the symmetric case, the DM anti-DM annihilation cross section will be suppressed by the small value of $r$ during CMB period.

After freeze-out, the evolution of $r$ is approximately described by following differential equation~\cite{Graesser:2011wi}: 
\begin{eqnarray}
\frac{d r }{dx} \simeq -   \frac{m_{\chi} M_{\text{Pl}} }{x^2}  \sqrt{\frac{\pi g_\ast }{45}} \left< \sigma_{\text{ann}} v \right>  Y_{\Delta\chi} \ r~,
\label{ratio2}
\end{eqnarray}
with $M_{\text{Pl}}$ the Planck scale, $x \equiv m_\chi / T$, $\left< \sigma_{\text{ann}} v \right>$ the thermally averaged annihilation cross-section, and $g_\ast$ the effective degree of freedom. 
In Fig.~\ref{fig:r_value} we present the value of $r$ during CMB period (labeled by $r_{\infty}$) as functions of $U(1)'$ coupling strength $g'$, for two benchmark spectrum favored by small-scale data. 
It shows that $r_{\infty}$ decrease rapidly as $g'$ growing. 

\begin{figure}
\begin{center}
\includegraphics[width=9cm]{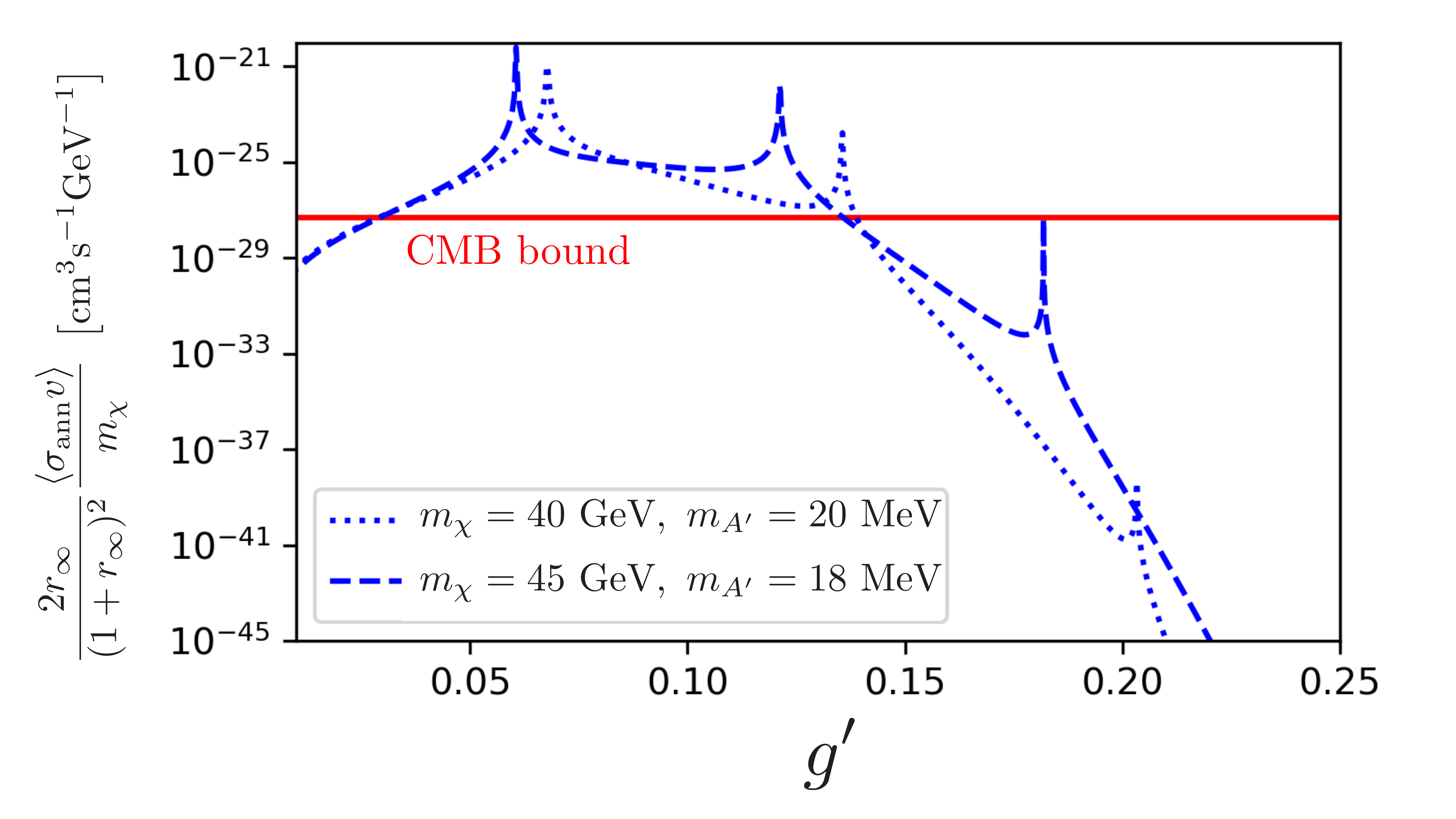}
\caption{Annihilation cross-section as functions of $g'$ for different benchmark spectrum. Region below red line is allowed by current CMB data.}
\label{fig:CMB}
\end{center}
\end{figure}

During the CMB period, DM become very non-relativistic and thus the non-perturbative effective, i.e. Sommerfeld enhancement, tends to be important. 
The annihilation cross-section during CMB period can be expressed as following~\cite{Cassel:2009wt}: 
\begin{eqnarray}
 (\sigma_{\text{ann}} v) = S(v) \times (\sigma_{\text{ann}} v)_0 ~,
\end{eqnarray}
with $(\sigma_{\text{ann}} v)_0  = { (\frac{3}{2}g')^4 }/{16 \pi m^2_{\chi}} $ the tree-level annihilation cross-section. 
Sommerfeld enhancement factor $S(v)$ is given by:
\begin{eqnarray}
S(v) = \frac{\pi}{\mathcal{A}(v)} \frac{\sinh(2\pi \mathcal{A}(v) \mathcal{B} )}{\cosh(2\pi \mathcal{A}(v) \mathcal{B}) - 
\cos(2\pi \sqrt{ \mathcal{B} - (\mathcal{A}(v) \mathcal{B})^2}) } ~,  \nonumber \\
\end{eqnarray}
with: 
\begin{eqnarray}
\mathcal{A}(v) = \frac{8\pi v}{9 g'^2}  \ , \
\mathcal{B} = \frac{27 g'^2 m_{\chi}}{8 \pi^3 m_{A'}}~.
\end{eqnarray}
It can be seen that the value of $S(v)$ can be very large when $v \ll 1$ and $ m_{A'} \ll m_{\chi} $.
This is why, for the symmetric DM scenario, the SIDM model with MeV scale mediator is highly constrained by CMB limit on DM annihilation cross-section~\cite{Bringmann:2016din}.

But for our asymmetric DM scenario, the situation is completely different. 
For the asymmetric DM scenario, annihilation can only happen between DM and anti-DM (which has a very small number density), and the annihilation cross-section changes as: 
\begin{eqnarray}
\left< \sigma_{\text{ann}} v \right>  \ \ \  &\Rightarrow& \ \ \ \left< \sigma_{\text{ann}} v \right> \times \frac{2 r_{\infty}}{(1 + r_{\infty} )^2} ~.\\\nonumber
 \text{symmetric case} & & \ \ \ \ \ \  \text{asymmetric case}
\end{eqnarray}
Thus the CMB limit on the asymmetric DM scenario is given by~\cite{Kawasaki:2021etm}: 
\begin{eqnarray}
\frac{2 r_{\infty}}{(1 + r_{\infty} )^2} \frac{\left< \sigma_{\text{ann}} v \right>}{m_\chi} \lesssim 5\times 10^{-28} \ \text{cm}^3\text{s}^{-1}\text{GeV}^{-1}~. \nonumber \\
\label{limit1}
\end{eqnarray}
It is easy to see that since the number density of anti-DM is much smaller than the number density of DM during CMB period (i.e. $r_{\infty}$ is very small), the limit on asymmetric DM model are greatly weakened.

In Fig.~\ref{fig:CMB} we present annihilation cross-section as functions of $g'$, for two benchmark spectrum favored by small-scale data. 
It shows that CMB limit can be easily escaped when $g' \gtrsim 0.20$. 
To solve the small-scale problems, $g'$ generally need to be much larger than 0.2, thus our model is safe from CMB bound.

\section{shortening the lifetime of $\phi$ by type-X 2HDM}\label{app:2HDM}

Our dark Higgs sector is charged under a $U(1)'$ and thus the lightest particle, i.e. $\phi$, can only decay to SM fermion pairs via coupling $\lambda S^\dagger S \Phi^\dagger \Phi$ ($\Phi$ is the SM Higgs doublet). 
However, due to the limit from Higgs invisible decay, parameter $\lambda$ can not be large enough to make the lifetime of $\phi$ to be shorter than 0.1 sec~\cite{Li:2023bxy}. 

To overcome this problem, here we consider a simple extension of the SM Higgs sector, the type-X two-Higgs-doublet-model (2HDM)~\cite{Aoki:2009ha}. 
The Higgs potential of type-X 2HDM with softly-broken $\mathbb{Z}_2$ symmetry is:  
\begin{eqnarray}
V_{2 \mathrm{HDM}} &&= -m_{11}^2 \Phi_1^{\dagger} \Phi_1-m_{22}^2 \Phi_2^{\dagger} \Phi_2-\left[m_{12}^2 \Phi_1^{\dagger} \Phi_2+\text { h.c. }\right]
\nonumber \\
&&+ \frac{1}{2} \lambda_1\left(\Phi_1^{\dagger} \Phi_1\right)^2+\frac{1}{2} \lambda_2\left(\Phi_2^{\dagger} \Phi_2\right)^2 \nonumber  \\
&& +\lambda_3\left(\Phi_1^{\dagger} \Phi_1\right)\left(\Phi_2^{\dagger} \Phi_2\right)  +\lambda_4\left(\Phi_1^{\dagger} \Phi_2\right)\left(\Phi_2^{\dagger} \Phi_1\right)  \nonumber \\ 
&& +\left[\frac{1}{2} \lambda_5\left(\Phi_1^{\dagger} \Phi_2\right)^2+ \text { h.c. }\right]~, 
\end{eqnarray}
where $\Phi_1$ and $\Phi_2$ are two Higgs doublets with the same quantum number. 
In type-X 2HDM, $\Phi_1$ only couples to leptons and $\Phi_2$ only couples to quarks. 
Yukawa couplings is given by: 
\begin{eqnarray}
\mathcal{L}_{\mathrm{Yukawa}} &&= - Y^u \bar{q}_L (i\sigma_2 \Phi_2^\ast ) u_R - Y^d \bar{q}_L \Phi_2 d_R \nonumber \\
&&- Y^e \bar{l}_L \Phi_1 e_R + \text {h.c. }~.
\end{eqnarray}

After electroweak symmetry breaking, these two Higgs doublets are expanded as: 
\begin{eqnarray}
&& \Phi_1 = \left(\begin{array}{c}
H^+_1 \\
\frac{1}{\sqrt{2}}(v_1 + h_1 + i a_1)
\end{array}\right) \ , \   \nonumber \\
&& \Phi_2 = \left(\begin{array}{c}
H^+_2 \\
\frac{1}{\sqrt{2}}(v_2 + h_2 + i a_2)
\end{array}\right)~. 
\end{eqnarray}
Parameter $\tan\beta$ is the ratio of two VEVs, $\tan\beta = v_2/v_1$.
The observed 125 GeV SM-like Higgs boson, to be labeled as $h$, is the mixing of $h_1$ and $h_2$: 
\begin{eqnarray}
\left(\begin{array}{c}
H \\
h
\end{array}\right)=\left(\begin{array}{cc}
\cos \alpha & \sin\alpha \\
-\sin\alpha & \cos\alpha
\end{array}\right)\left(\begin{array}{l}
h_1 \\
h_2
\end{array}\right)~,
\end{eqnarray}
with $H$ the CP-even Higgs in addition to $h$. 

To be consistent with current measurement of 125 GeV Higgs, we consider the so-called alignment limit where $\sin(\beta-\alpha)=1$~\cite{Bernon:2015qea}.
In the alignment limit, it is convenient to consider the Higgs basis, where: 
\begin{eqnarray}
\left(\begin{array}{c}
\Phi_H \\
\Phi_h
\end{array}\right)=\left(\begin{array}{cc}
\sin\beta & -\cos\beta \\
\cos\beta & \sin\beta
\end{array}\right)\left(\begin{array}{l}
\Phi_1 \\
\Phi_2
\end{array}\right)~,
\end{eqnarray}
where $\Phi_h$ can be considered as the SM doublet which takes responsibility for electroweak symmetry breaking, and $\Phi_H$ will have no VEV.
The coupling between $h$ and SM fermions are the same as their SM value and thus to be consistent with current Higgs data. 
While the coupling between $H$ and fermions depend on $\tan\beta$:
\begin{eqnarray}
\mathcal{L}^{\mathrm{even}}_{\mathrm{Yukawa}} &&= -\frac{m_u}{v_{EW}} \left( h \bar{u}u - \frac{1}{\tan\beta} H \bar{u}u \right)  \nonumber \\
&&-\frac{m_d}{v_{EW}} \left( h \bar{d}d - \frac{1}{\tan\beta} H \bar{d}d \right)  \nonumber \\
&& -\frac{m_l}{v_{EW}} \left( h \bar{l}l + {\tan\beta} H \bar{l}l \right)~. 
\end{eqnarray}
So it is possible to make MeV scale dark Higgs $\phi$ short-lived via the the mixing between $\phi$ and $H$, provided $\tan\beta$ is large enough. 

Next, we consider the portal terms between dark sector and visible sector: 
\begin{eqnarray}
V_{\mathrm{portal}} &&= \bar{\lambda}_1  S^\dagger S \left( \Phi^\dagger_h \Phi_H + \text { h.c. }  \right) 
+ \bar{\lambda}_2  S^\dagger S \left( \Phi^\dagger_h \Phi_h \right)   \nonumber \\
&& + \bar{\lambda}_3  S^\dagger S \left( \Phi^\dagger_H \Phi_H  \right)~, 
\end{eqnarray}
where we assume $\bar{\lambda}_2$ and $\bar{\lambda}_3$ to be negligible small. 
After electroweak and $U(1)'$ symmetry breaking, mixing term $\bar{\lambda}_1 v_s v_{EW} \phi H$ is induced. 
Because $H$ is much heavier than $\phi$, thus the mixing angle $\theta$ is approximately given by: 
\begin{eqnarray}
\theta \approx  \bar{\lambda}_1 \frac{ v_s v_{EW} }{ m^2_H }~,
\end{eqnarray}
so the Yukawa coupling between $s$ and electron pair is: 
\begin{eqnarray}
Y_{\phi e} \approx \bar{\lambda}_1 \frac{ v_s v_{EW} }{ m^2_H } \frac{m_e}{v_{EW}} \tan\beta ~,
\end{eqnarray}

To get rid of the limit from $\Delta N_\text{eff}$ from BBN, we need the thermalization of visible sector and dark sector~\cite{Ibe:2021fed} at least at a temperature $T= m_\phi$, which requires
\begin{eqnarray}
\Gamma= Y_{\phi e}^2 \frac{m_\phi}{8\pi^2} >  H(T=m_\phi)= 1.66 \sqrt{g} \frac{T^2}{m_{pl}}~,
\end{eqnarray}
which set a limit that, 
\begin{eqnarray}
\bar{\lambda}_1 \frac{ v_s v_{EW} }{ m^2_H } \tan\beta > 1.5\times 10^{-4}~.
\end{eqnarray}
Note that this term could be taken as an effective mixing $\sin \theta$ in the scenario which $s$ mixes with the standard model Higgs. From the Fig.12 of~\cite{Ibe:2021fed}, it clearly shows that $\sin \theta = 10^{-4}$ can not make the dark sector thermalize with visable sector but pretty close, therefore our calculation is consistent with the result with~ \cite{Ibe:2021fed}. The obstacle of thermalzation in only SM Higgs scenario mainly from the limit of Higgs invisible decay which sets $\sin \theta < 10^{-5}$. In our case, the new Higgs boson only has very weak coupling with quarks, then its production rate is very small at the LHC, therefore current limit on it is very weak and it is even allowed to be low as tens of GeV~\cite{Chun:2021rtk}. As an conservative estimation, we can set $m_H=200$ GeV, $\bar{\lambda}_1 \sim 1 $, with $v_s \sim 20$ MeV,  we find $\tan \beta > 1.5$ 
enough to satisfy above limit. A larger $\tan \beta$ can further enhance such coupling.

On the other hand, the lifetime of $\phi$ is: 
\begin{eqnarray}
\tau_{\phi} = \frac{1}{\Gamma} \approx \frac{8 \pi^2 m_H^4 }{\bar{\lambda}_1^2 v_s^2 
 m_e^2 \tan^2\beta m_\phi }~.
\end{eqnarray}
When $m_H=200$ GeV, $\bar{\lambda}_1 \sim 1 $, $v_s \sim 20$ MeV, and $m_\phi \sim 4$ MeV, the lifetime of $\phi$ is about $\frac{0.2}{\tan^2\beta}$ sec. 
So it is easy to make $\phi$ short-lived. 

We note that the introduction of the additional Higgs would loop-induced kinetic mixing of the visible photon and dark photon, which is highly constrained by dark matter direct detection. However, it has been found that the loop induced kinetic mixing of the visible photon and dark photon is highly suppressed~\cite{Koren:2019iuv}, which appears at five loops in our model. The combination of loop factors and coupling suppression provides sufficient suppression to maintain consistency with current limits.


\bibliography{ref}
\bibliographystyle{apsrev4-1}


\end{document}